\documentclass[pra,aps,nofootinbib,twocolumn]{revtex4-2}

\usepackage{epsfig,amssymb,amsmath}

\usepackage{color}
\usepackage[utf8]{inputenc}
\usepackage{amsmath, amssymb, mathtools, braket, graphicx, float, bbold, bm, ulem}
\usepackage{hyperref}

\hypersetup{
    colorlinks=true,
    linkcolor=blue
}

\usepackage[colorinlistoftodos]{todonotes}
\usepackage{soul}
\usepackage{appendix}

\def\>{\rangle}\def\<{\langle}
\def\togli#1{}

\newcommand{\Mod}[1]{\left|#1\right|}

\usepackage{listings}
\begin{document}


\title{Error-analysis for the Sorkin and Peres tests performed on a quantum computer}

   
\author{Simanraj Sadana$^1$, Lorenzo Maccone$^2$, Urbasi Sinha$^{3}$ }
\email{usinha@rri.res.in}
\affiliation{ 
\vbox{$^1$INFN
    Sez.~Pavia, University~of Pavia, via Bassi 6, I-27100 Pavia,
    Italy}
\vbox{$^2$Dip.~Fisica and INFN
    Sez.~Pavia, University~of Pavia, via Bassi 6, I-27100 Pavia,
    Italy}
\vbox{$^3$Light and Matter Physics, Raman Research Institute, Bengaluru-560080, India}
}
\begin{abstract}
We use quantum computers to test the foundations of quantum mechanics through quantum algorithms that implement some of the experimental tests as the basis of the theory's postulates. These algorithms can be used as a test of the physical theory under the premise of a perfect hardware or as a test of the hardware under the premise that quantum theory is correct. In this paper, we show how the algorithms can be used to test the efficacy of a quantum computer in obeying the postulates of quantum mechanics. We study the effect of different types of errors on the results of experimental tests of the postulates. A salient feature of this error analysis is that it is deeply rooted in the fundamentals of quantum mechanics as it highlights how systematic errors affect the quantumness of the quantum computer.

\end{abstract}
\maketitle
\section{Introduction}
Quantum computers can serve as test-beds for the laws of quantum mechanics. Two such experimental tests: the Sorkin test for Born's rule \cite{sorkin, caslav, zykowski, lee, urbasi, laflamme, rengaraj, usinha3} and the Peres test to verify the \emph{complex} nature of quantum mechanics \cite{perescompl, stu,toni,walther,gregor,janwei, werner} were performed on a quantum computer in \cite{sadana2022peres}. This was achieved by constructing quantum circuits that implement both the tests and running them on a noisy intermediate-scale quantum (NISQ) device (in \cite{sadana2022peres}, the hardware was provided by Rigetti \cite{rigettiweb}). However, owing to noises in a quantum computer, the aforementioned tests generally yield significant deviations from the expected values. Taking these systematic errors into account bridges the gap between theoretical expectation and experimental observation, thus confirming the correctness of quantum theory.

Conversely, with the premise that quantum mechanics is correct, the deviations in the test-results may be used to get some information about the efficacy of a quantum computer in simulating quantum mechanics. In this paper, we study the variation of the results of the Peres and Sorkin tests with respect to different types of noises present in a quantum computer. The significance of such a study is that it highlights which aspect (or postulate(s)) of quantum mechanics is effectively violated in a quantum computer when certain kinds of errors are present. Such an analysis tells us whether the quantum computer virtually behaves as if quantum mechanics is quaternionic or whether the superposition principle and/or Born’s rule (measurement probability is square-modulus of scalar product of the wave-function with some basis vector) are notionally violated due to hardware faults. As these tests are based on the postulates of quantum mechanics, we are effectively testing the quantumness of the hardware.

The approach of this study is to include artificial noise-models in the quantum circuit and by changing the amount of error(s) present, observe the variation in the results of the Peres/Sorkin test. The variation of the results of the Peres test indicates how effectively the quantum computer obeys the superposition principle and is \emph{complex} in nature. On the other hand, the variations in the result of the Sorkin test tells us the degree to which Born's rule is satisfied. 
However, an important point to note here is that the test results depend on the initial state used and therefore, the efficacy of a quantum computer also depends on the states that are in use.

The paper is organised as follows. We review the Peres and Sorkin tests using quantum computers with ideal circuits. Then we artificially introduce different types of noises in the circuit and run the tests to plot the variation of the test-results as a function of the amount of error present in the circuit. The tests are performed for a set of random input states (see appendix \ref{app:randStates} on how the random states are prepared) to highlight the fact that the effect of noise is dependent on the input state.

\section{Review of Peres and Sorkin test on quantum computers}\label{sec:review}
In \cite{sadana2022peres} the Peres and Sorkin tests were performed in two different ways: (1) Separate circuits for Peres and Sorkin tests and (2) One circuit to perform a joint Peres-Sorkin test. In this paper, we consider only the joint test without loss of generality. The simplest (and with least number of gates in the circuit) way to perform the joint test is to prepare a random state
\begin{align}
    \ket{\psi} = a_{0}\ket{00} + a_{1}\ket{01} + a_{2}\ket{10} 
    \label{eq:initState}
\end{align}
and get the projections $\braket{\Phi_{012} | \psi}$, $\braket{\Phi_{i,i\oplus_31} | \psi}$ for $i \in \{0,1,2\}$ and  $\braket{B(i) | \psi}$ for $i \in \{0,1,2\}$, where $B(i)$ is binary equivalent of the decimal number $i$ and 
\begin{align}
    \label{eq:triProjection}
    \ket{\Phi_{012}} =& \frac{\ket{00} + \ket{01} + \ket{10}}{\sqrt{3}} \\
    \label{eq:biProjection}
    \ket{\Phi_{ij}} =& \frac{\ket{B(i)} + \ket{B(j)}}{\sqrt{2}}
\end{align}
which can be obtained by appropriate unitary transformations at the end of the circuit followed by measurement in the computational basis.

For the Peres test, the required quantities are 
\begin{align}
    \gamma_{ij} =& \frac{\left|\braket{\Phi_{ij} | \psi}\right|^2 - \left|\braket{B(i)|\psi}\right|^2 - \left|\braket{B(j)|\psi}\right|^2}{2 \left|\braket{B(i)|\psi}\right| \left|\braket{B(j)|\psi}\right|}, \\
    F =& \gamma_{01}^2 + \gamma_{12}^2 + \gamma_{20}^2 - 2\gamma_{01}\gamma_{12}\gamma_{20}
\end{align}
where, $F=1$ ensures the sufficiency of complex numbers for quantum mechanics \cite{perescompl}. To test Born's rule, we need the Sorkin's parameter \cite{sorkin} defined as
\begin{align}
    \kappa =& \left|\braket{\Phi_{012} | \psi}\right|^2 - 2\left(\left|\braket{\Phi_{01} | \psi}\right|^2 + \left|\braket{\Phi_{12} | \psi}\right|^2 + \left|\braket{\Phi_{20} | \psi}\right|^2\right) \nonumber \\
    & + \left|\braket{B(0) | \psi}\right|^2 + \left|\braket{B(1) | \psi}\right|^2 + \left|\braket{B(2) | \psi}\right|^2
\end{align}
where $\kappa=0$ implies that Born's rule is correct. In fact, this method can be extended to higher number of qubits to check higher-order Sorkin parameters (see appendix \ref{app:kappaN}).
In \cite{sadana2022peres} we showed that on a quantum computer the values of $F$ and $\kappa$ were significantly different from their expected values if quantum mechanics was complex and Born's rule was correct. However, taking into account the systematic errors of the machine into the theoretical simulations, the observed results matched with the expected values. Therefore, the conclusion was that indeed the postulates under consideration are correct. On the other hand, if we assume \emph{a priori} that the postulates of quantum mechanics are correct, then these deviations can be used to gain information about ``how quantum is a quantum computer". Therefore, we systematically include noise-models in the quantum circuit that performs the joint Peres-Sorkin test and see its response. 

We use IBM's Qiskit platform to study the variation of the results of Peres and Sorkin tests with noise. In Qiskit, the errors added to the noise model have their instruction appended to the gates in the circuit. In general, we can consider a composite noise model containing noises of different types like readout error in measurements, depolarizing error and thermal-relaxation error. For say, $N$ different types of noise, let $\{p_1, p_2, \cdots, p_N\}$ be the set of parameters that quantify the amount of the errors present. Then we can plot the quantities $F(\{p_i\})$ and $\kappa(\{p_i\})$ with respect to different amounts of these errors.


\subsection{Circuit}
\begin{figure}
    \centering
    \includegraphics[width=1\hsize]{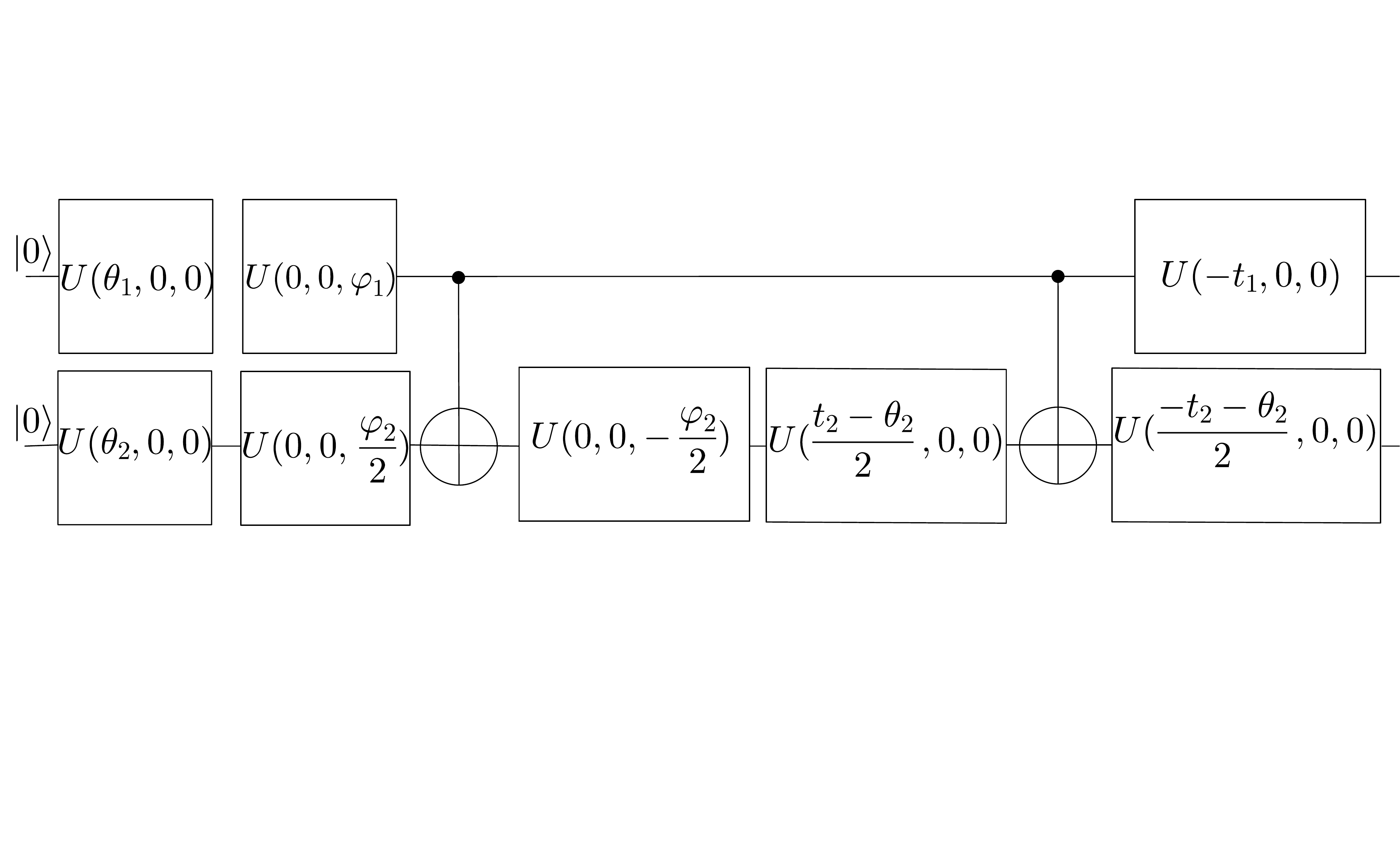}
    \caption{The quantum circuit for joint Peres and Sorkin tests. The parameters $\theta_1$, $\theta_2$, $\varphi_1$ and $\varphi_2$ are used to generate the intitial state of the form in Eq.~\eqref{eq:initStateMod}. On the other hand, $t_1$ and $t_2$ are used to choose the projection of the chosen initial state.}
    \label{fig:mainCirc}
\end{figure}

The quantum circuit to perform the joint Sorkin-Peres test prepares an initial state of the form
\begin{align}
    \label{eq:initStateMod}
    \ket{\psi} =& \cos\frac{\theta_1}{2} \ket{00} + \mathrm{e}^{\mathrm{i}\varphi_1}\sin\frac{\theta_1}{2}\cos\frac{\theta_2}{2}\ket{01} \nonumber \\
    &+ \mathrm{e}^{\mathrm{i}(\varphi_1+ \varphi_2)}\sin\frac{\theta_1}{2}\sin\frac{\theta_2}{2}\ket{11}
\end{align}
which can be constructed from the state $\ket{00}$ using the following operations,
\begin{align}
    \ket{\psi} = CU_{01}(0,0,\varphi_2)CU_{01}(\theta_2,0,0)U(0,0,\varphi_1)U(\theta_1, 0, 0)\ket{00}
\end{align}
where
\begin{align}
\label{eq:1qubitU}
    U(\theta, \varphi, \lambda) =& \left(\begin{matrix}
                                        \cos\frac{\theta}{2} & - \mathrm{e}^{\mathrm{i}\lambda}\sin\frac{\theta}{2} \\
                                        \mathrm{e}^{\mathrm{i}\varphi}\sin\frac{\theta}{2} & \mathrm{e}^{\mathrm{i}(\varphi + \lambda)}\cos\frac{\theta}{2}
                                    \end{matrix}\right)
\end{align}
Following the preparation stage the state is projected onto a state of the form,
\begin{align}
    \ket{\phi} =& \cos\frac{t_1}{2} \ket{00} + \sin\frac{t_1}{2}\cos\frac{t}{2}\ket{01} + \sin\frac{t_1}{2}\sin\frac{t_2}{2}\ket{11}
\end{align}
The simplified quantum circuit that achieves this is shown in figure \ref{fig:mainCirc}. Note that the state in Eq.~\eqref{eq:initStateMod} is different from the one in Eq.~\eqref{eq:initState} as instead of $\ket{10}$ we have $\ket{11}$. We use the latter only because it requires less number of operations/gates. However, as the only requirement for the test is to have a 3-level state, this difference is not important. We include noise in this ideal circuit using the noise-models in the qiskit package.

We run the (noisy) circuit for a set of random input states (see appendix for the method of preparing random states). For every value of the noise parameter the circuit is run for $10^5$ shots. The plots for random input states highlights the fact that the effect of noise on the variation of the values of $\kappa$ and $F$ depends on the input state, but belong to the same family of curves for a particular type of noise. To check the statistical fluctuation of the results (because of limited sample-size), we run the tests for a specific state and repeat the tests $30$ times to get a sample which is then bootstrapped \cite{bootcite} to generate a $99\%$ confidence interval (see appendix \ref{app:statFluc}).

\section{Noises}\label{sec:Noises}
\subsection{Readout noise}
\begin{figure}
    \centering
    \includegraphics[width=1\hsize]{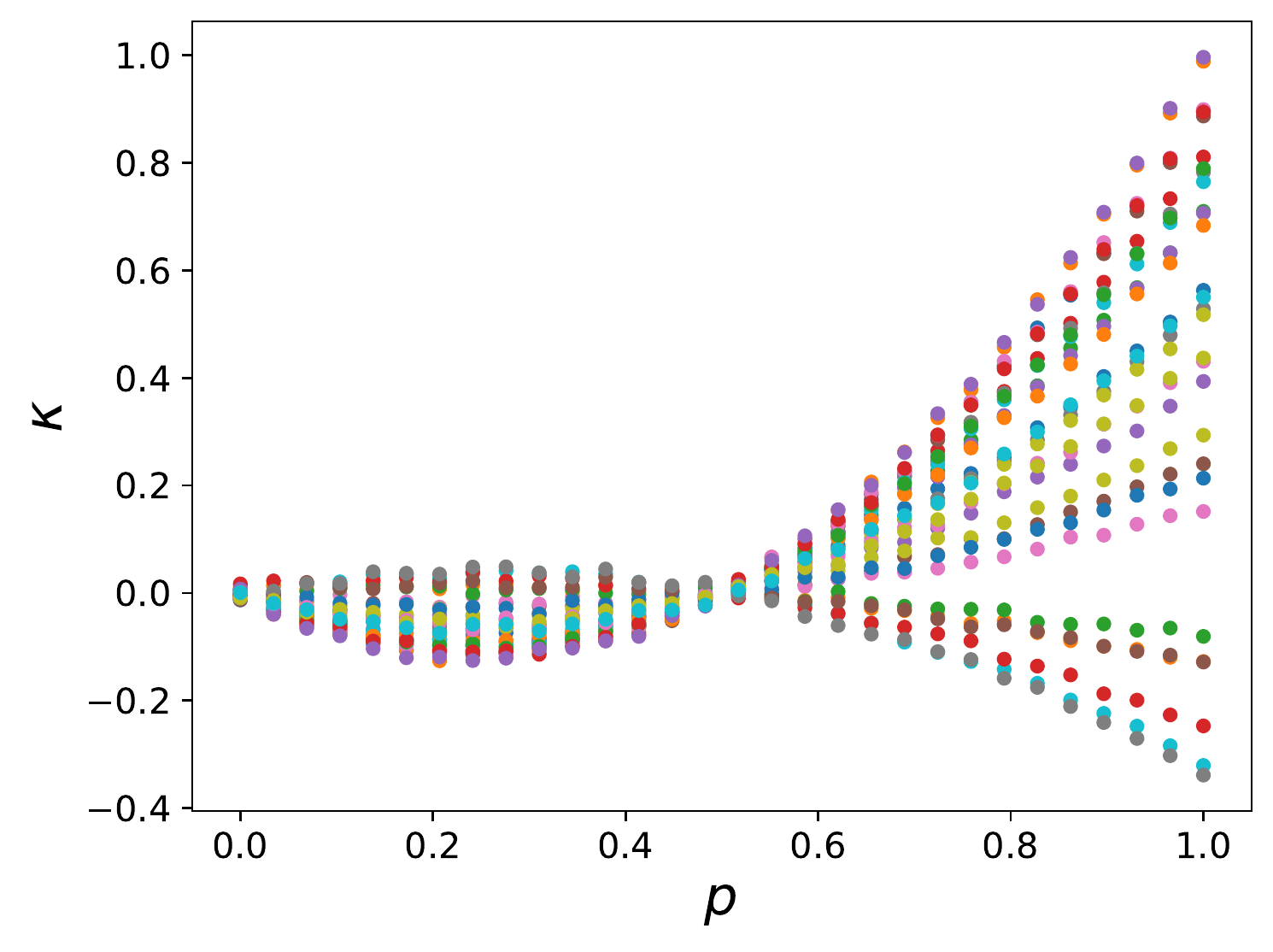}
    \caption{The variation of $\kappa$ in the presence of different amounts of readout error simulated for a set of random input states (different coloured dots correspond to different states). A common feature of all the curves is that they are all quadratic in nature. Another notable feature is that $\kappa = 0$ for readout errors of $p=0$ and $p=0.5$ irrespective of the initial state.}
    \label{fig:figKReadoutRandom}
\end{figure}

\begin{figure}
    \centering
    \includegraphics[width=1\hsize]{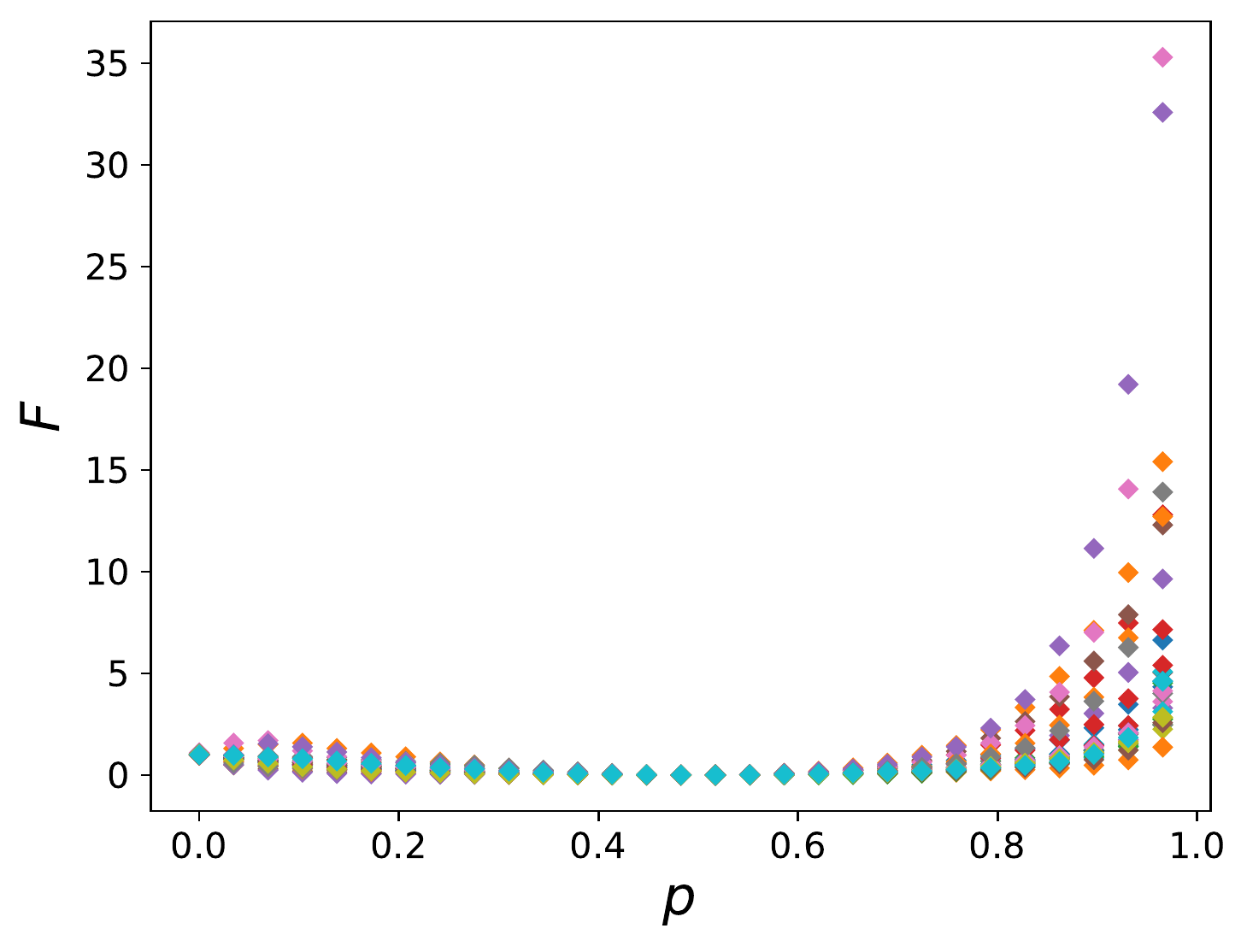}
    \caption{The variation of $F$ in the presence of different amounts of readout error simulated for a set of random input states (different coloured dots correspond to different states). A common feature of all the curves is that there is a threshold of the amount readout error beyond which the superposition principle is violated. Below that threshold, the quantum computer behaves as if quantum mechanics is quaternionic.}
    \label{fig:figFReadoutRandom}
\end{figure}

Readout noise is the probabilistic bit-flip at the measurement stage of the circuit. For the purpose of this paper, 
we assume that the readout-error is symmetric, i.e., the probabilities of flips $0 \rightarrow 1$ and $1 \rightarrow 0$ are equal, say $p$. However, in general, the readout error can be asymmetric depending on the hardware. For example, if the experiment is photonic, a photon can be missed due to the dead-time of a detector and on the other hand, dark-counts can lead to false clicks. In general, these two processes will happen with different degrees leading to an asymmetric readout error. In Qiskit, the readout-error is represented as the matrix,
\begin{align}
    P =& \left(\begin{matrix}1-p & p \\ p & 1-p\end{matrix}\right)
\end{align}
The noise-model is constructed and added to the circuit on both the qubits.
The noise-model is parameterized by $p$ which can be varied over the interval $[0,1]$. The result of simulation of the noisy circuit with random input states yields the result shown in figures \ref{fig:figKReadoutRandom} and \ref{fig:figFReadoutRandom}.

\begin{figure}
    \centering
    \includegraphics[width=1\hsize]{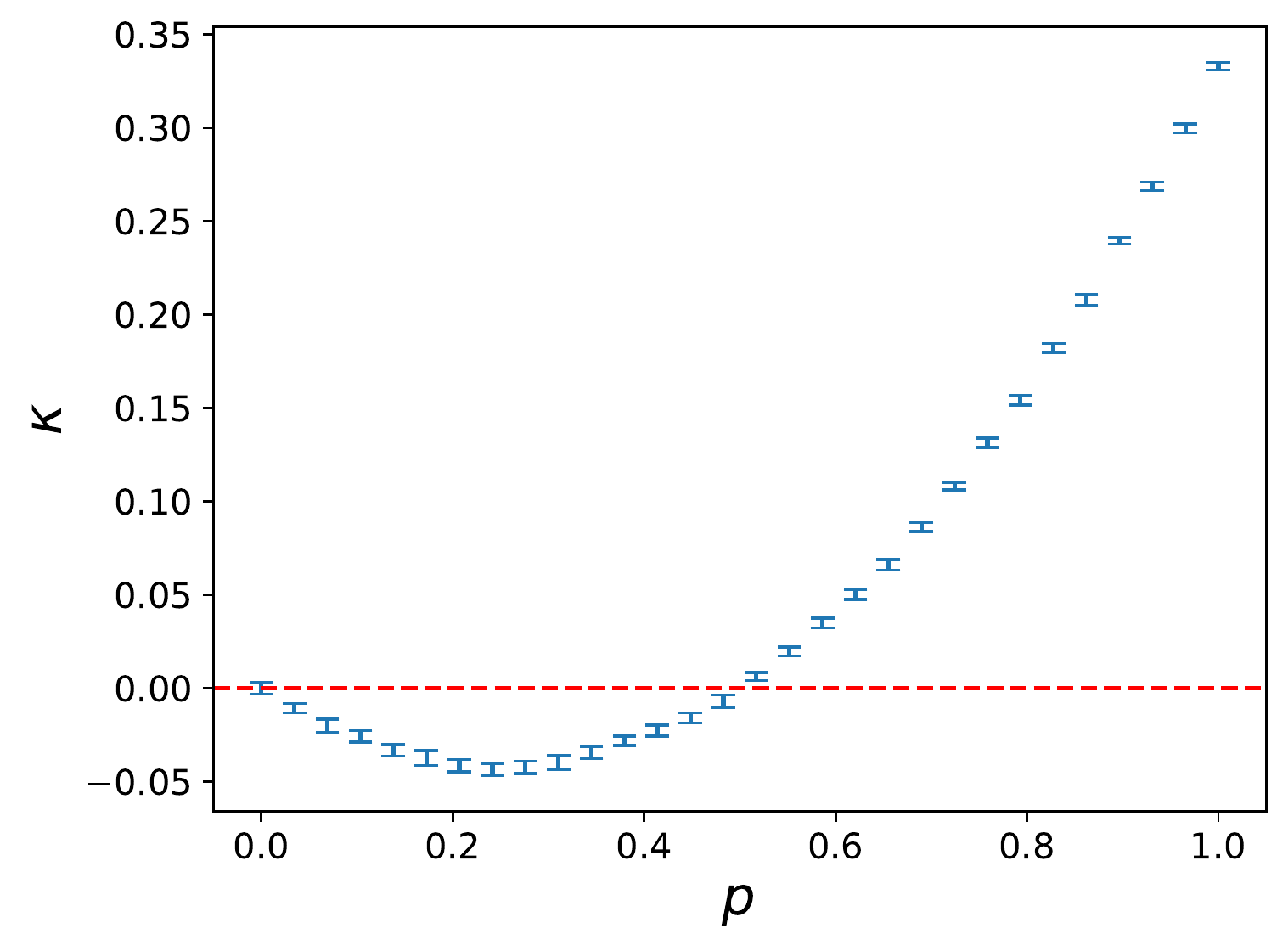}
    \caption{The variation of $\kappa$ in the presence of different amounts of readout error simulated for the input state given in Eq.~\eqref{eq:specificState}.}
    \label{fig:FigKReadoutOne}
\end{figure}

\begin{figure}
    \centering
    \includegraphics[width=1\hsize]{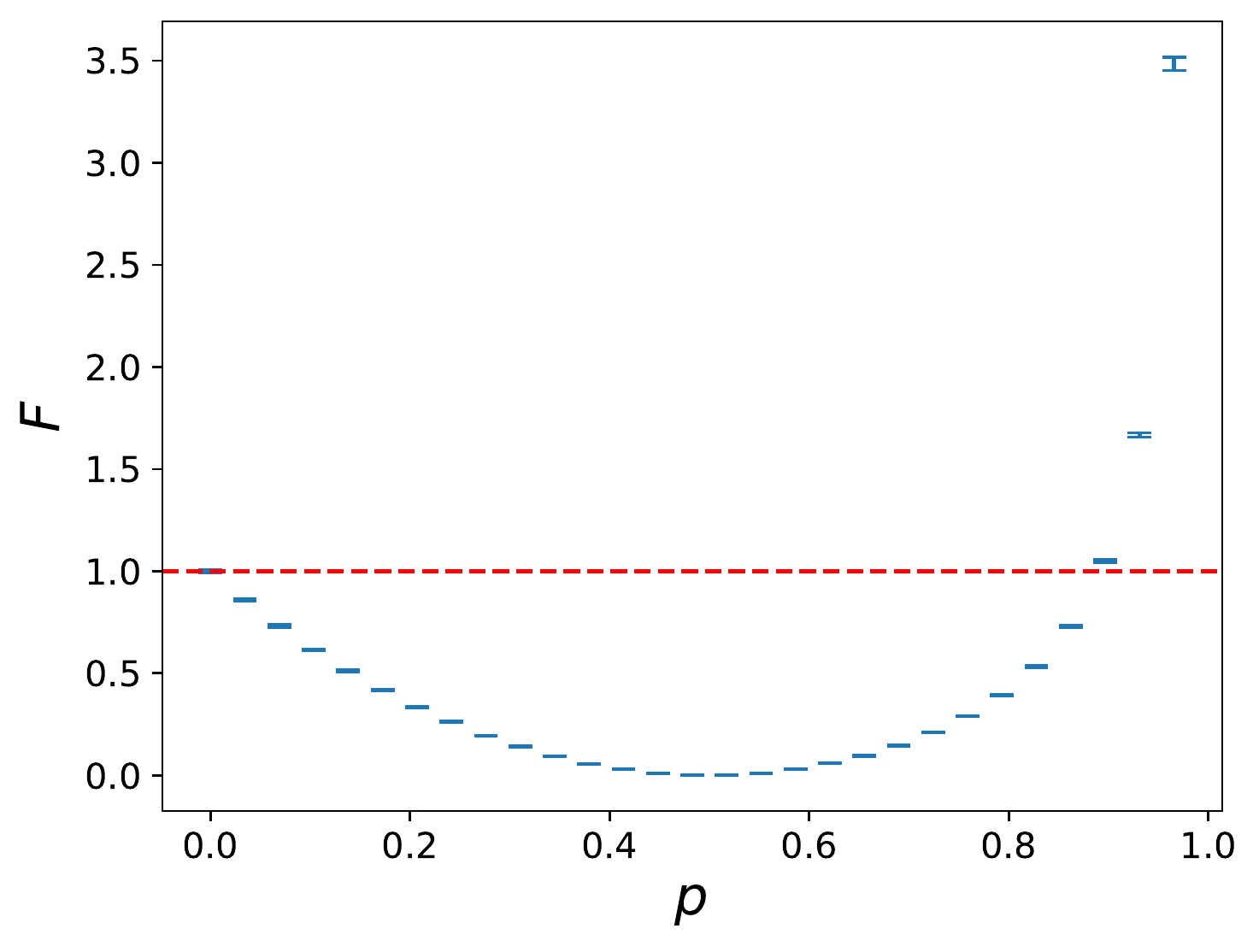}
    \caption{The variation of $F$ in the presence of different amounts of readout error simulated for a set of random input states. In this case, the threshold beyond which the superposition principle is violated is $p \approx 0.9$. Below this threshold, the quantum computer virtually behaves as if quantum mechanics is quaternionic.}
    \label{fig:FigFReadoutOne}
\end{figure}

To closely analyze the effect of readout error we simulated the noisy circuit with a specific input state, i.e., 
\begin{align}
    \label{eq:specificState}
    \ket{\psi} =& \frac{1}{\sqrt{3}} \left(\ket{00} + \mathrm{e}^{\mathrm{i}\pi/4}\ket{01} + \mathrm{e}^{\mathrm{i}\pi/2}\ket{11}\right)
\end{align}
The result of the simulation are shown in figure \ref{fig:FigKReadoutOne} and \ref{fig:FigFReadoutOne}. A key feature of the result is that it highlights the range of readout error for which the superposition principle is effectively broken. In this case, a readout error of more than approximately $0.9$ leads to the effective breakdown of the superposition principle. For readout errors less than that threshold, the superposition principle holds but the quantum computer behaves as if quantum mechanics is quaternionic. Especially in the case of $p=0.5$, Sorkin's parameter is zero implying that Born's rule holds and yet $|F| < 1$ implies that the quaternionic behaviour of the quantum computer is not because of the breakdown of Born's rule. However, in practical situations, the readout error is not so big that a violation of the superposition occurs. From the graphs in figure \ref{fig:figFReadoutRandom} it is evident that for small amounts of readout error, the superposition principle stays intact.
\subsection{Depolarizing noise}
Depolarizing noise turns a state into a mixture of that state and a maximally-mixed state,
\begin{align}
    \rho \xrightarrow{\text{Depolarizing noise}} (1-p) \rho + p \frac{\mathbb{1}}{2^n}
\end{align}
where $n$ is the number of qubits. In qiskit, depolarizing noise is included in the circuit by making the constituent gates noisy. The noise-model is constructed and added to the one-qubit $U(\theta, \varphi, \lambda)$ operation and the two-qubit CNOT gates that constitute the circuit in figure \ref{fig:mainCirc}.
%
The parameter $p$ is the parameter for the amount of depolarizing noise present. 

\begin{figure}
    \centering
    \includegraphics[width=1\hsize]{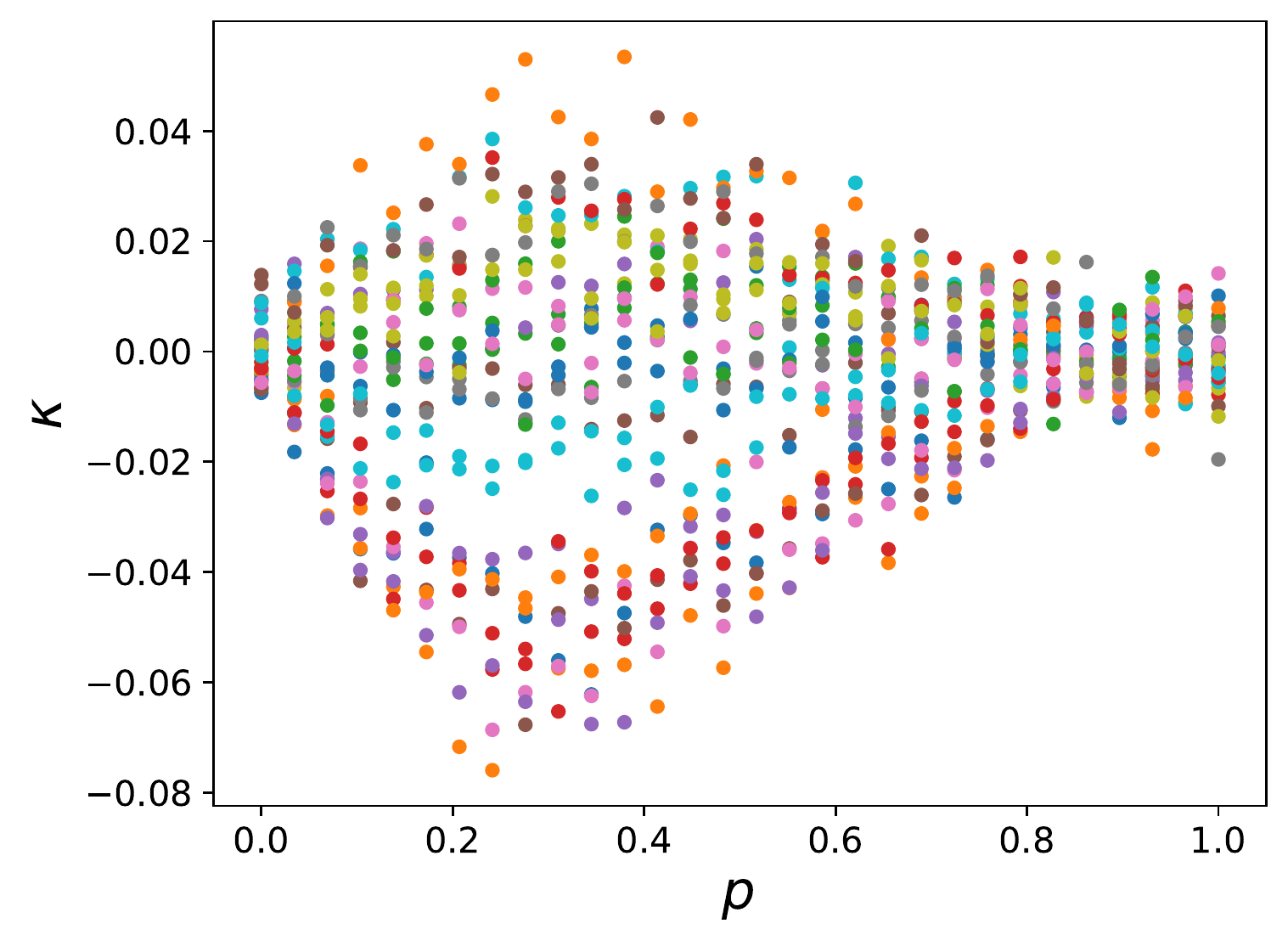}
    \caption{Variation of $\kappa$ with different amounts of depolarizing noise for a set of random states (different coloured dots represent different states).}
    \label{fig:figKDepolarizingRandom}
\end{figure}

\begin{figure}
    \centering
    \includegraphics[width=1\hsize]{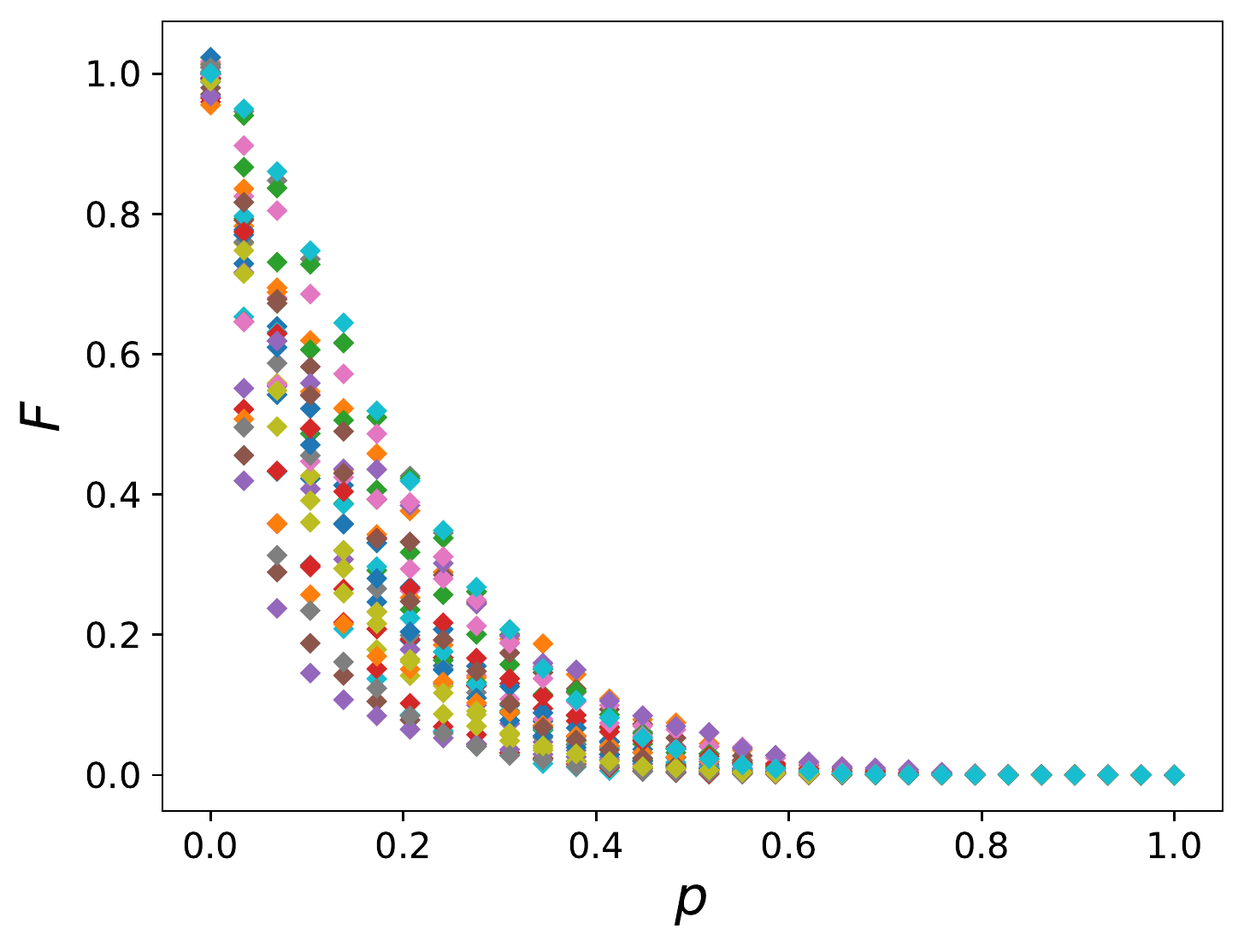}
    \caption{Variation of $F$ with different amounts of depolarizing noise for a set of random states (different coloured dots represent different states). Depolarizing noise does not cause a violation of the superposition. However, it does make the effective behaviour of the quantum computer quaternionic.}
    \label{fig:figFDepolarizingRandom}
\end{figure}

\begin{figure}
    \centering
    \includegraphics[width=1\hsize]{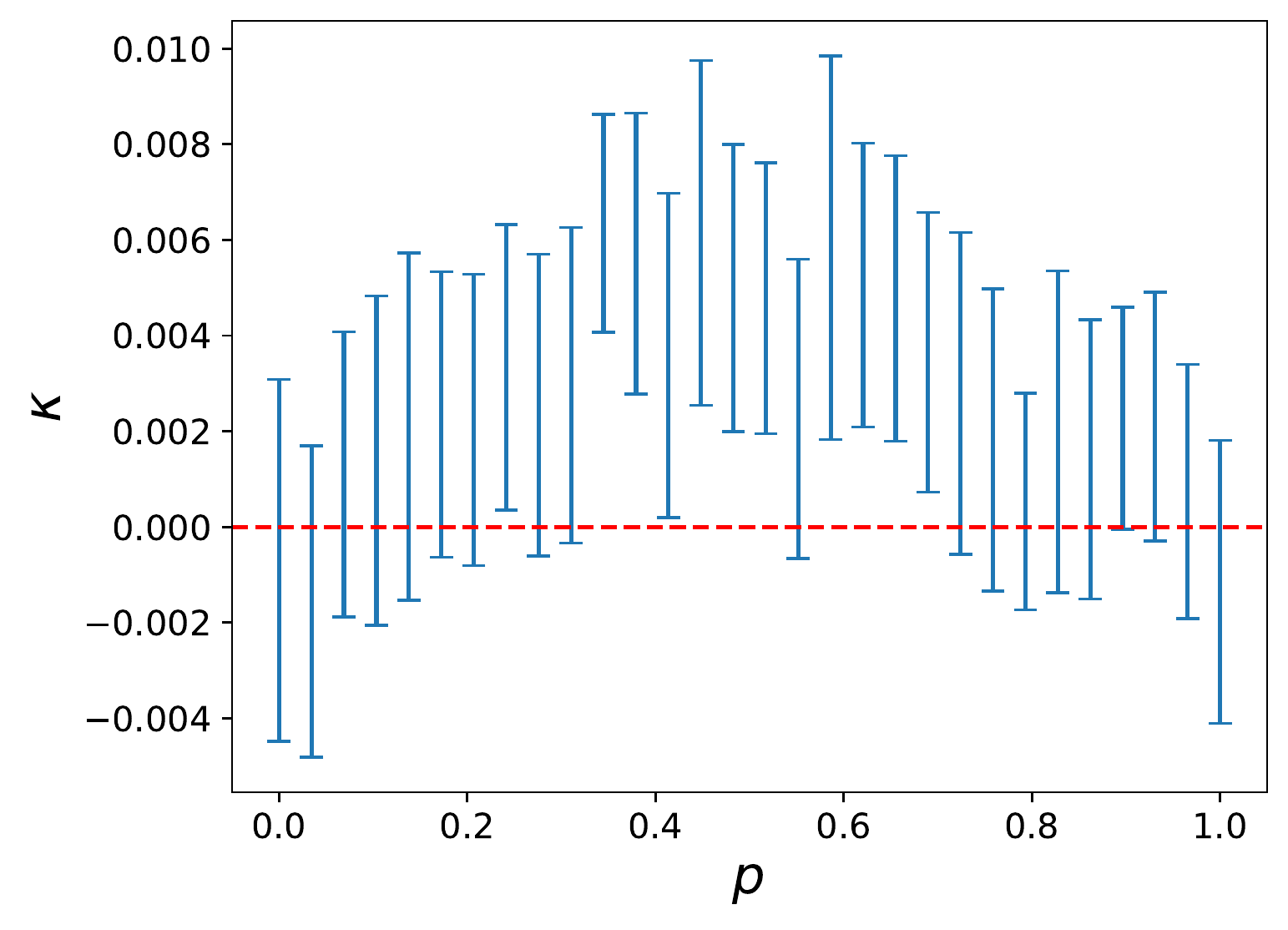}
    \caption{Variation of $\kappa$ with different amounts of depolarizing noise for the specific state in Eq.~\eqref{eq:specificState}. As discussed in the main text, the value of $\kappa$ is zero at both extremes. For $p=0$ the system is ideal so Born's rule holds and for $p=1$, the state is a maximally-mixed state for which $\kappa$ becomes zero.}
    \label{fig:figKDepolarizingOne}
\end{figure}

\begin{figure}
    \centering
    \includegraphics[width=1\hsize]{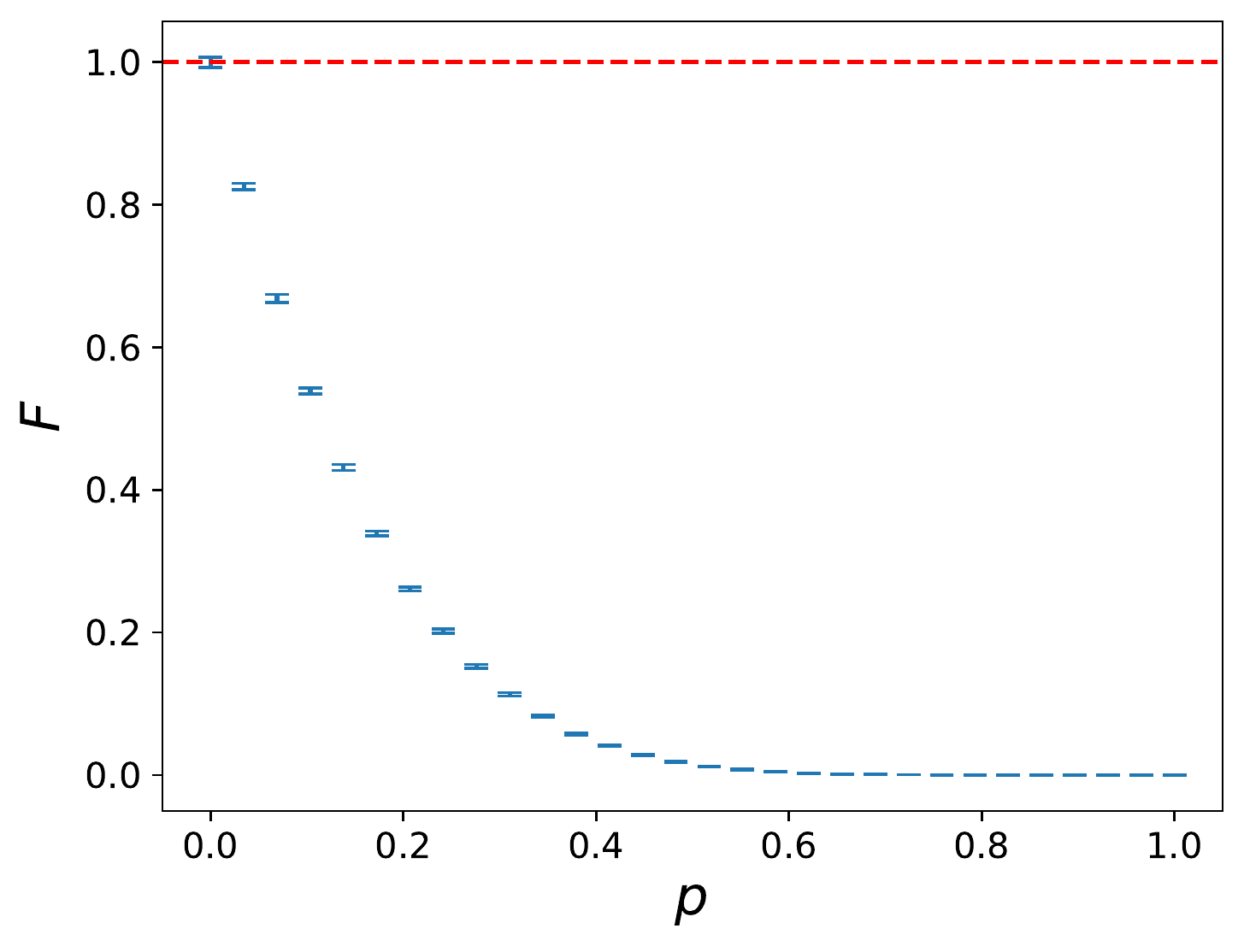}
    \caption{Variation of $F$ with different amounts of depolarizing noise for the specific state in Eq.~\eqref{eq:specificState}. Depolarizing noise does not cause any violation of the superposition principle, but only makes the quantum computer effectively quaternionic.}
    \label{fig:figFDepolarizingOne}
\end{figure}

\begin{figure}
    \centering
    \includegraphics[width=1\hsize]{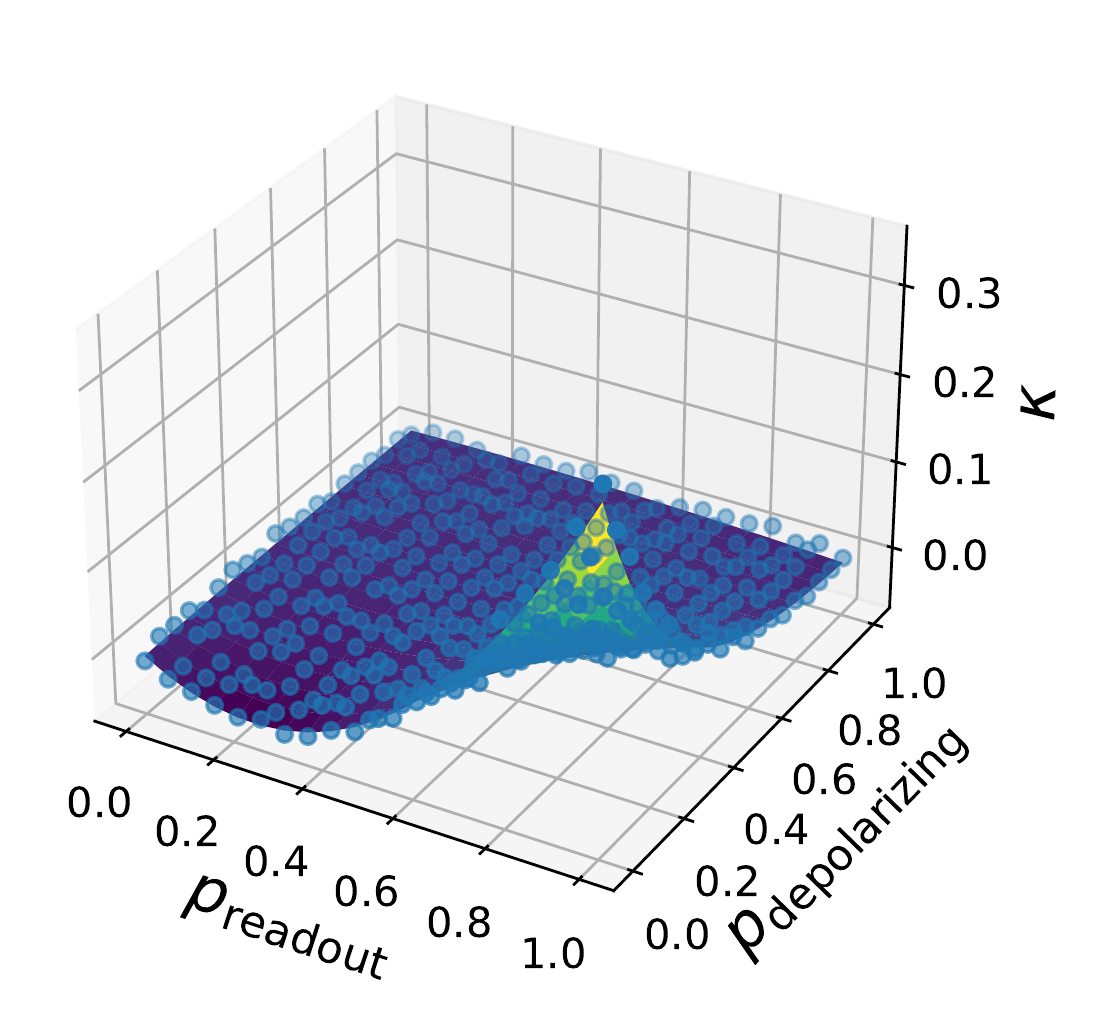}
    \caption{Variation of $\kappa$ with a combination of readout and depolarizing error for input state in Eq.~\eqref{eq:specificState}.}
    \label{fig:figKReadoutDepol}
\end{figure}

\begin{figure}
    \centering
    \includegraphics[width=1\hsize]{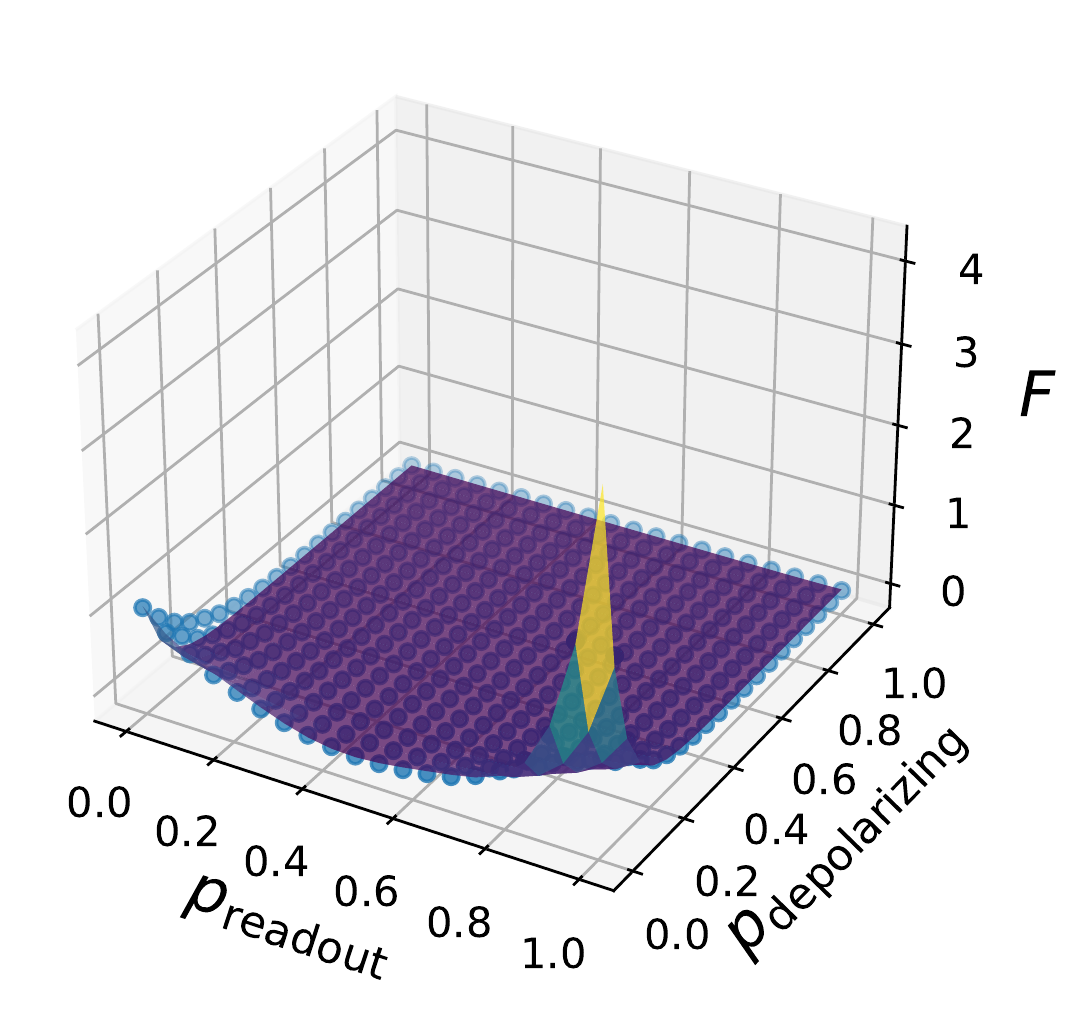}
    \caption{Variation of $F$ with a combination of readout and depolarizing error for input state in Eq.~\eqref{eq:specificState}.}
    \label{fig:figFReadoutDepol}
\end{figure}

Figures \ref{fig:figKDepolarizingRandom} and \ref{fig:figFDepolarizingRandom} show the results with random input states. Figures \ref{fig:figKDepolarizingOne} and \ref{fig:figFDepolarizingOne} show the graphs for the specific state in Eq.~\eqref{eq:specificState}. The result of the noisy Sorkin test shows that the deviation from Born's rule initially increases with the increase in depolarizing noise, but after a point the value of $\kappa$ decrease back to zero. This happens because $\kappa$ for the maximally-mixed state is zero and therefore, as the proportion of the maximally-mixed state dominates, the value of $\kappa$ gets closer to zero. The same can be seen in the case of the specific state (defined in Eq.~\eqref{eq:specificState}) as shown in figure \ref{fig:figKDepolarizingOne}. Unlike the case of readout error, depolarizing noise does not lead to the effective failure of the superposition principle. However, when depolarizing noise is present with another type of noise, it can lead to the failure of superposition principle. As an example, figures \ref{fig:figKReadoutDepol} and \ref{fig:figFReadoutDepol} show the results of the tests when the readout and depolarizing noises are present together. 


\subsection{Thermal-relaxation noise}
Thermal-relaxation noise is parameterized by two time-scales, wiz., $T_1$ which quantifies the time it takes for a qubit to decay from $\ket{1}$ to $\ket{0}$, and $T_2$ which is the coherence-time of the qubit. To include this noise in the quantum circuit, we need to specify the gate-times of the various operations involved. The relevant operations in the circuit in figure \ref{fig:mainCirc} are the 1-qubit unitary operation defined in Eq.~\eqref{eq:1qubitU}, CNOT, reset and measure with operation-times $100$ ns, $300$ ns, $1000$ ns and $1000$ ns respectively. And then the noise-model is constructed given the average $T_1$ and $T_2$ which are then sampled from a Gaussian-distribution with the same mean values. Then we add the thermal noise to the corresponding gates in the circuit.

\begin{figure}
    \centering
    \includegraphics[width=1\hsize]{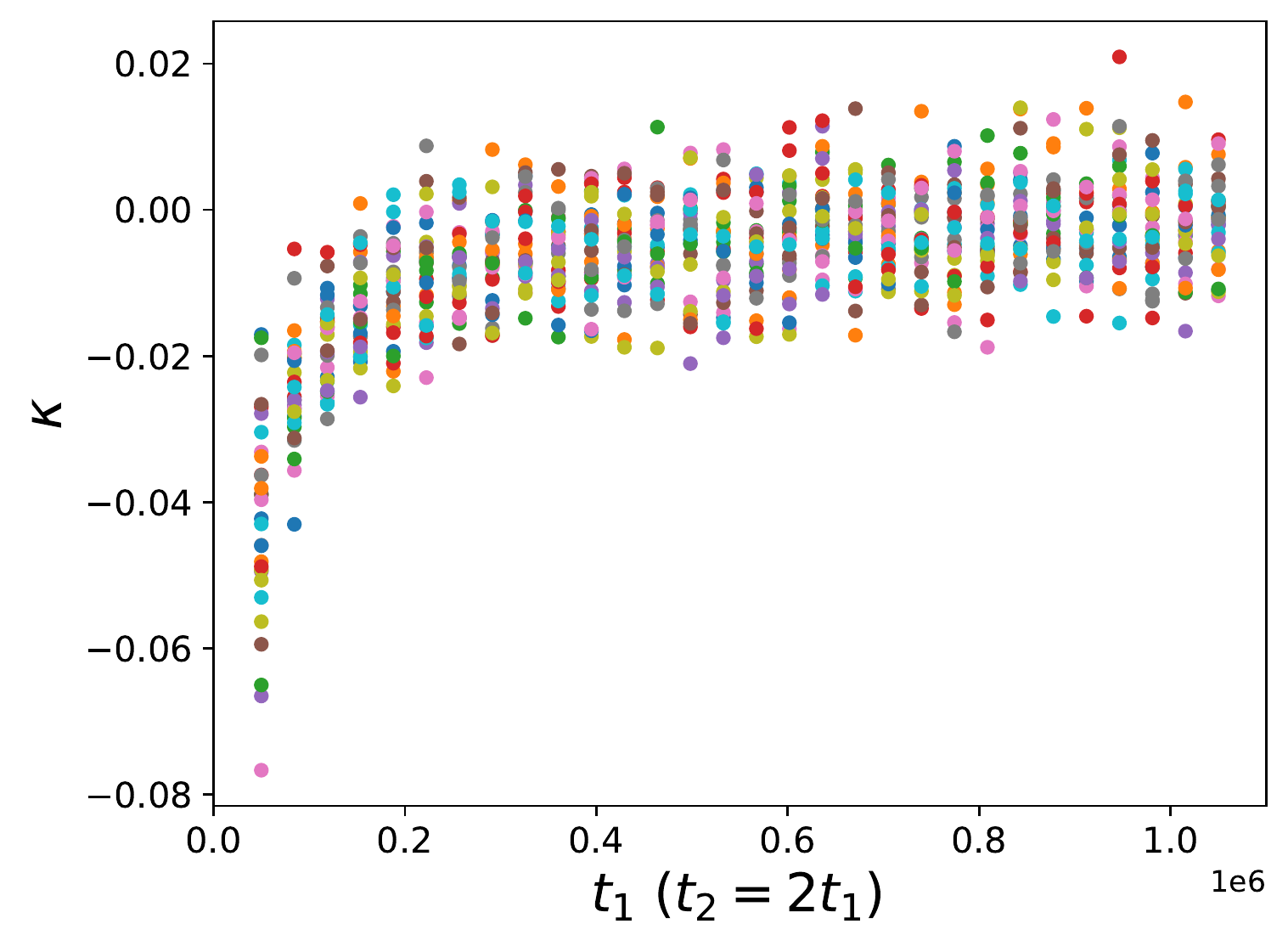}
    \caption{Variation of $\kappa$ with thermal-relaxation noise for which $T_2 = 2T_1$. Born's rule is violated for values of $T_1$ less than the gate-times. But as $T_1$ becomes larger than all the gate-times, the value of $\kappa$ approaches the ideal value of zero.}
    \label{fig:figKThermalRandom}
\end{figure}
\begin{figure}
    \centering
    \includegraphics[width=1\hsize]{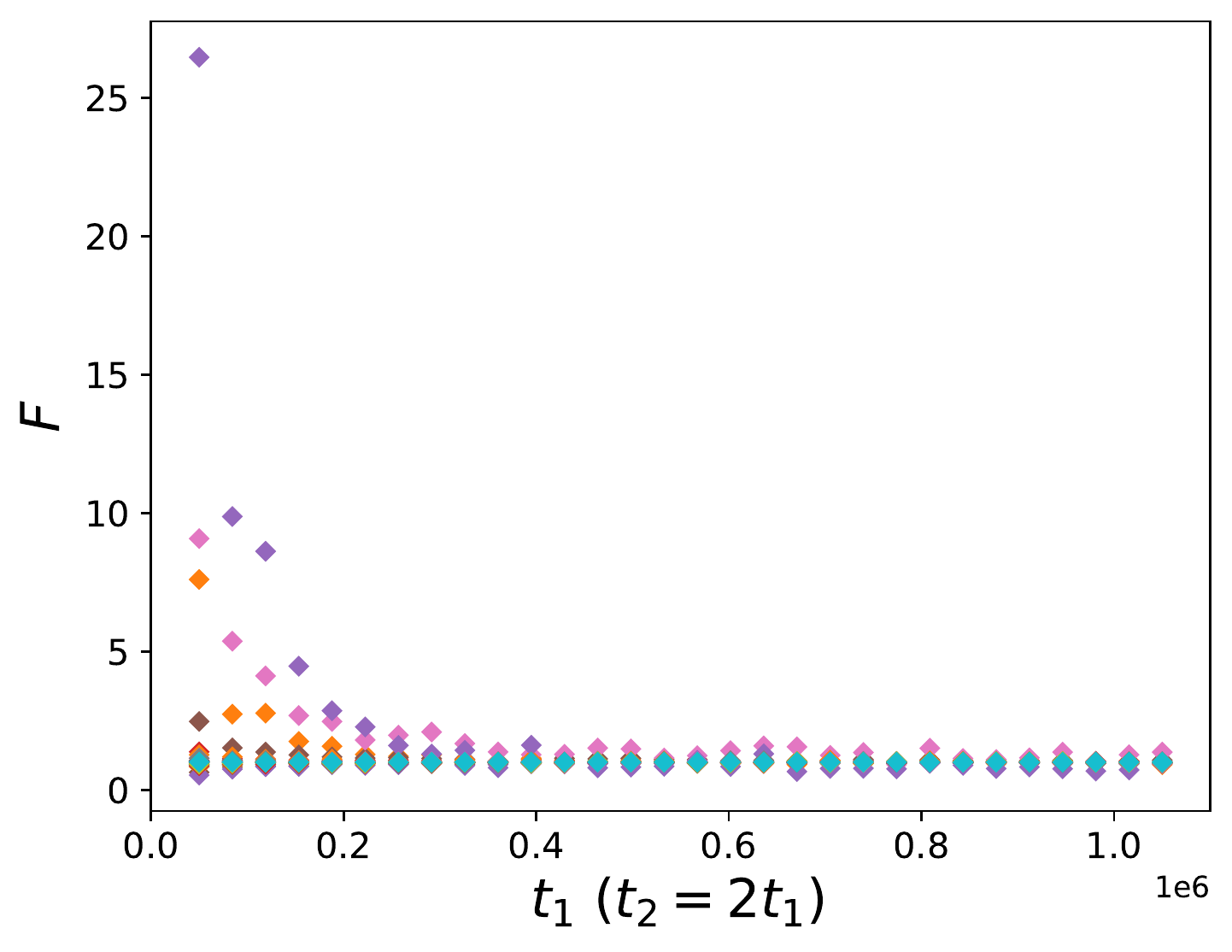}
    \caption{Variation of $F$ with thermal-relaxation noise for which $T_2 = 2T_1$. The plots for a set of random states (different colours correspond to different states) show that depending on the initial state, the superposition principle may be violated. But as $T_1$ becomes larger than all the gate-times, the value of $F$ approaches the ideal value of unity.}
    \label{fig:figFThermalRandom}
\end{figure}
\begin{figure}
    \centering
    \includegraphics[width=1\hsize]{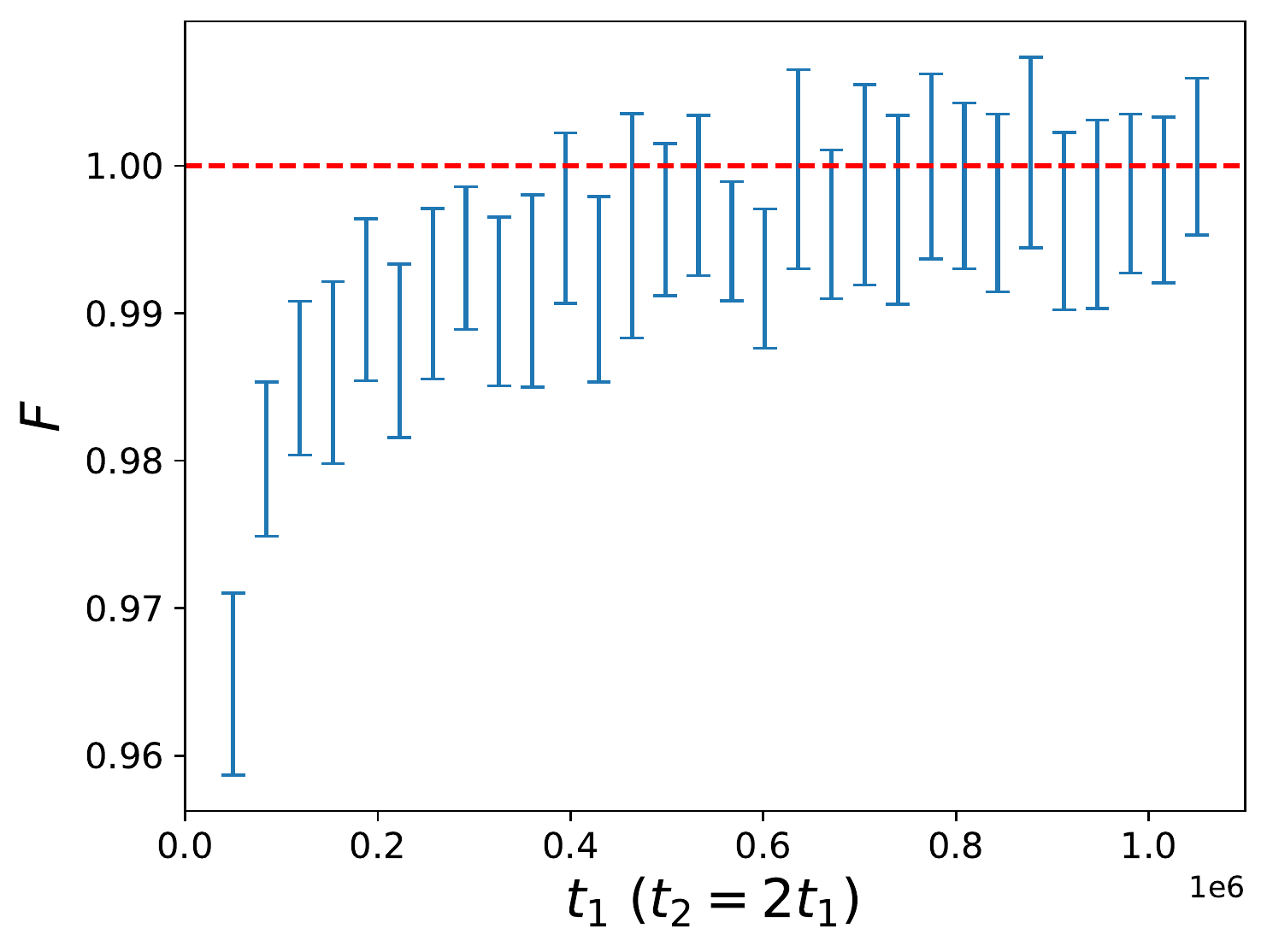}
    \caption{Variation of $\kappa$ with thermal-relaxation noise with $T_2 = 2T_1$. The result is for the specific state in Eq.~\eqref{eq:specificState}}
    \label{fig:FigFThermalOne}
\end{figure}

The constraint on the values of the times is $T_2 \leq 2 T_1$. For the purpose of an example, we set $T_2 = 2 T_1$ and see the variation of $\kappa$ and $F$ with changing $T_1$, as shown in figures \ref{fig:figKThermalRandom} and \ref{fig:figFThermalRandom}. As the $T_1$ time becomes equal to or more than the largest instruction-time, the values of $\kappa$ and $F$ approach the ideal values. However, whether the superposition is valid depends on the initial state, as for some state the value of $F$ crosses one and for others it stays below one. For the specific state in Eq.~\eqref{eq:specificState}, the superposition principle is always obeyed as shown in figure \ref{fig:FigFThermalOne}.
\begin{figure}
    \centering
    \includegraphics[width=1\hsize]{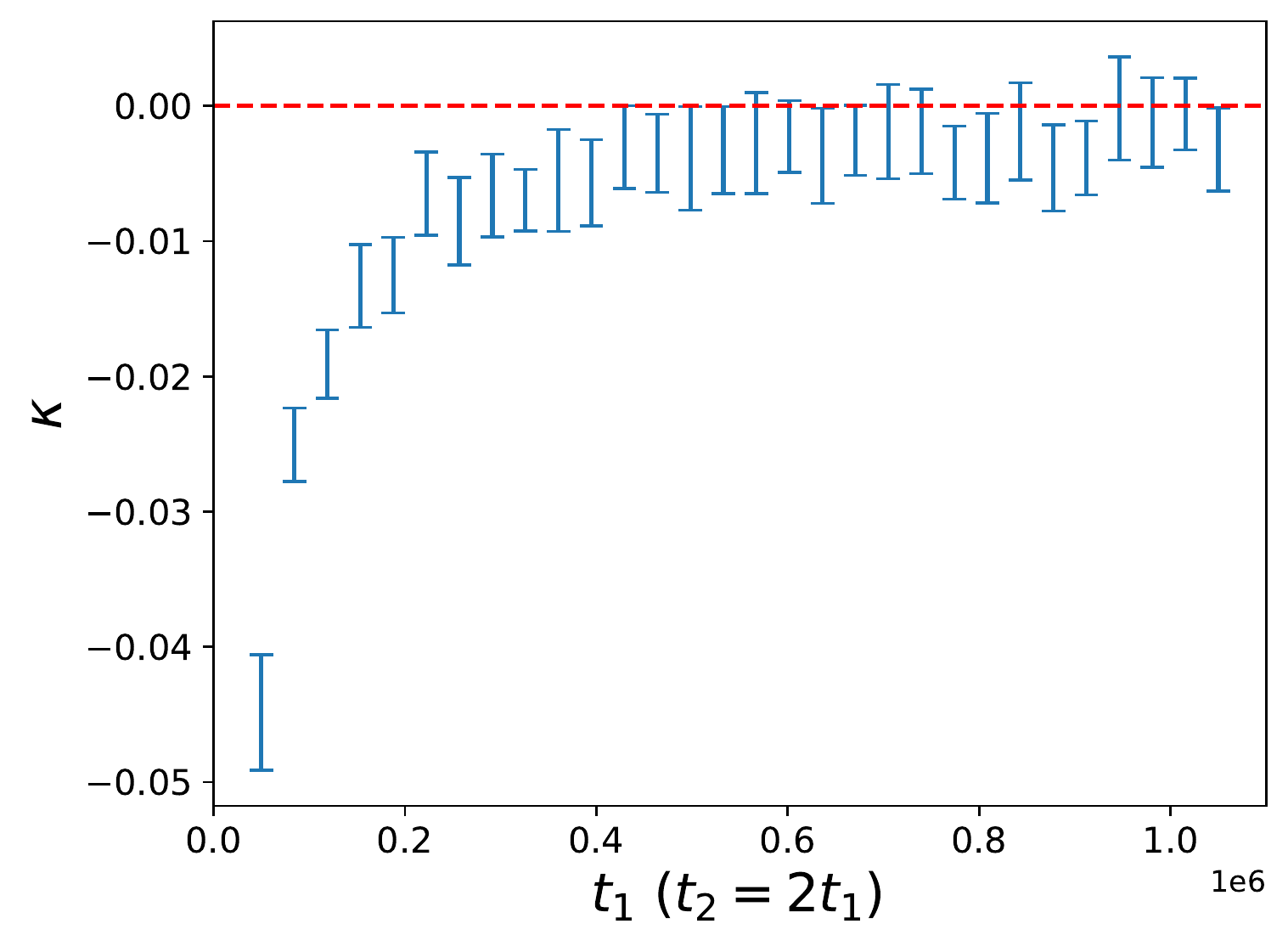}
    \caption{Variation of $\kappa$ with thermal-relaxation noise with $T_2 = 2T_1$. The result is for the specific state in Eq.~\eqref{eq:specificState}.}
    \label{fig:FigKThermalOne}
\end{figure}

\section{Conclusion}
If the postulates of quantum mechanics are assumed to be true, then the postulates themselves can be used to check the efficacy of a quantum computer as the working of a quantum computer is based on the quantum mechanical postulates. In this paper, we studied the variation of the results of Peres and Sorkin's test on a quantum computer and used the deviation as an evidence for systematic errors in the quantum computer. Moreover, it is possible to identify which postulate of quantum mechanics is effectively not followed by the quantum computer because of the noises. The complex/quaternion nature of quantum mechanics is verified by using the Peres test while Born's rule is checked by using the result of Sorkin's test. As such an error analysis is based on the postulates of quantum mechanics, they can serve as deep quantum benchmarks for the quantum computer. 

This material is based upon work supported by the U.S. Department of Energy, Office of Science, National Quantum Information Science Research Centers, Superconducting Quantum Materials and Systems Center (SQMS) under contract number DE-AC02-07CH11359. US acknowledges partial support provided by the Ministry of Electronics and Information Technology
(MeitY), Government of India under grant for Centre for Excellence in Quantum Technologies with Ref. No.
4(7)/2020 - ITEA, QuEST-DST project Q-97 of the Govt. of India and the QuEST-ISRO research grant.

\appendixpage
\appendix
\section{Generating random states}\label{app:randStates}
The values of parameter $\theta$ and $\varphi$ for the preparation of random states are generated as follows. We take a random variable $r$ uniformly distributed over the interval $[0,1]$ transform it appropriately to get the required probability distributions for $\theta$ and $\varphi$. Let the required functions be $\theta(r)$ and $\varphi(r)$. Then
\begin{align}
    P(\theta) d\theta =& P(r) dr \nonumber\\
    \implies~ \int\limits_{0}^{\theta} P(\theta) d\theta =& \int\limits_{0}^{r} P(r) dr \nonumber\\
    \implies~ \int\limits_{0}^{\theta} \frac{\sin\theta}{2} d\theta =& \int\limits_{0}^{r} 1 dr \nonumber\\
    \implies~ \frac{1 - \cos\theta}{2} =& r \nonumber\\
    \implies~ \theta(r) =& \cos^{-1}\left(1 - 2r\right)
\end{align}
Now we can generate a random number $r$ uniformly over $[0.1]$ and calculate $\theta(r)$ with the required distribution. Similarly,
\begin{align}
    P(\varphi) d\varphi =& P(r) dr \nonumber\\
    \implies~ \int\limits_{0}^{\varphi} P(\varphi) d\varphi =& \int\limits_{0}^{r} P(r) dr \nonumber\\
    \implies~ \int\limits_{0}^{\varphi} \frac{1}{2\pi} d\varphi =& \int\limits_{0}^{r} 1 dr \nonumber\\
    \implies~ \frac{\varphi}{2\pi} =& r \nonumber\\
    \implies~ \varphi(r) =& 2 \pi r
\end{align}

\section{Higher-dimensional Sorkin test} \label{app:kappaN}
We prove by mathematical induction. Sorkin's parameter with 3 terms,
\begin{align}
    |x_1 + x_2 + x_3|^2 =& |x_1 + x_2|^2 + |x_1 + x_3|^2 + |x_2 + x_3|^2 \nonumber\\
    & -|x_1|^2 - |x_2|^2 - |x_3|^2
\end{align}

If the 3-term expression is true as above then with 4 terms,
\begin{align}
|x_1+x_2+x_3+x_4|^2 =& |x_1+x_2|^2 + |x_1+x_3|^2 + |x_1+x_4|^2\nonumber\\
&+|x_2+x_3|^2 + |x_2+x_4|^2\nonumber\\
&+|x_3+x_4|^2\nonumber\\
& -2|x_1|^2 - 2|x_2|^2 - 2|x_3|^2 - 2|x_4|^2
\end{align}

If the 4-term expression is true as above then with 5 terms,
\begin{align}
    |x_1+x_2+x_3+x_4+x_5|^2 =& |x_1+x_2|^2+|x_1+x_3|^2+|x_1+x_4|^2\nonumber \\
    &+|x_1+x_5|^2 +|x_2+x_3|^2+|x_2+x_4|^2 \nonumber \\
    &+|x_2+x_5|+|x_3+x_4|^2+|x_3+x_5|^2\nonumber\\ 
&+ |x_4+x_5|^2-3|x_1|^2-3|x_2|^2-3|x_3|^2\nonumber \\
&-3|x_4|^2-3|x_5|^2
\end{align}

Observing the trend, we can guess that for $n$ terms,
\begin{align}
\label{eq:kappaForn}
    \left|\sum\limits_{i=1}^{n}x_i\right|^2 = \frac{1}{2}\sum\limits_{i,j}^{n,n}|x_i+x_j|^2 - (n-1)\sum\limits_{i}^{n}|x_i|^2
\end{align}

The general expression can be proven by mathematical induction.

For $n=1$,

LHS = $|x_1|^2$

RHS = $|x_1|^2$

Therefore, it holds for $n=1$. We assume that the expression is true for $n=k$. Now, for $n=k+1$,

\begin{align}
    \left|\sum\limits_{i=1}^{k} x_i + x_{k+1}\right|^2 =& \left|\sum\limits_{i=1}^{k}x_i\right|^2 + |x_{k+1}|^2 + \sum\limits_{i=1}^{k}2\Re\{x^*_i x_{k+1}\} \nonumber \\
    =& \left|\sum\limits_{i=1}^{k}x_i\right|^2 + |x_{k+1}|^2 + \sum\limits_{i=1}^{k}\left(|x_i + x_{k+1}|^2 \right.\nonumber \\
    &\left.- |x_i|^2 - |x_{k+1}|^2\right)
\end{align}

Using equation \eqref{eq:kappaForn}:
\begin{widetext}
\begin{align}
\label{eq:kappaFornplus1}
    \left|\sum\limits_{i=1}^{k} x_i + x_{k+1}\right|^2 =& \frac{1}{2}\sum\limits_{i,j}^{k,k}|x_i+x_j|^2 - (k-1)\sum\limits_{i}^{k}|x_i|^2 + |x_{k+1}|^2 + \sum\limits_{i=1}^{k}\left(|x_i + x_{k+1}|^2 - |x_i|^2 - |x_{k+1}|^2\right)\nonumber \\
    =& \frac{1}{2}\sum\limits_{i,j}^{k,k}|x_i+x_j|^2 + \sum\limits_{i=1}^{k}|x_i + x_{k+1}|^2 + |x_{n+1}|^2 - k\sum\limits_{i}^{k}|x_i|^2 -k |x_{k+1}|^2\nonumber \\
\end{align}
\end{widetext}
Now observing the fact that
\begin{widetext}
\begin{align}
    \frac{1}{2}\sum\limits_{i,j}^{k+1,k+1}|x_i+x_j|^2 =& \frac{1}{2}\sum\limits_{i,j}^{k,k+1}|x_i+x_j|^2 + \frac{1}{2}\sum\limits_{j}^{k+1}|x_{k+1}+x_j|^2 \nonumber\\ 
    =& \frac{1}{2}\sum\limits_{i,j}^{k,k}|x_i+x_j|^2 + \frac{1}{2}\sum\limits_{i}^{k}|x_i+x_{k+1}|^2 + \frac{1}{2}\sum\limits_{j}^{k+1}|x_{k+1}+x_j|^2 \nonumber\\ 
    =& \frac{1}{2}\sum\limits_{i,j}^{k,k}|x_i+x_j|^2 + \frac{1}{2}\sum\limits_{i}^{k}|x_i+x_{k+1}|^2 + \frac{1}{2}\sum\limits_{j}^{k}|x_{k+1}+x_j|^2  + |x_{k+1}|^2\nonumber\\ 
    =& \frac{1}{2}\sum\limits_{i,j}^{k,k}|x_i+x_j|^2 + \sum\limits_{i=1}^{k}|x_i + x_{k+1}|^2 + |x_{k+1}|^2
\end{align}
\end{widetext}
Equation \eqref{eq:kappaFornplus1} simplifies to
\begin{align}
    \left|\sum\limits_{i=1}^{k} x_i + x_{k+1}\right|^2 =& \frac{1}{2}\sum\limits_{i,j}^{k+1,k+1}|x_i+x_j|^2 - k\sum\limits_{i}^{k+1}|x_i|^2 \nonumber \\
\end{align}

Therefore, by mathematical induction, equation \eqref{eq:kappaForn} is true for all $n\geq 1$.

With this generalized expression, we can define higher-dimensional Sorkin's parameter
\begin{align}
    \kappa_n =& \left|\sum\limits_{i=1}^{n}x_i\right|^2 - \frac{1}{2}\sum\limits_{i,j}^{n,n}|x_i+x_j|^2 + (n-1)\sum\limits_{i}^{n}|x_i|^2
\end{align}

which can be used to test Born's rule in higher-dimensional systems. Re-writing it in a form suitable for programming,
\begin{align}
    \kappa_n =& \left|\sum\limits_{i=1}^{n}x_i\right|^2 - \sum\limits_{i,j>i}^{n-1,n}|x_i+x_j|^2 + (n-2)\sum\limits_{i}^{n}|x_i|^2
\end{align}
\section{Statistical fluctuations} \label{app:statFluc}
The quantities that are considered here depend on the probabilities of measurement outcomes. Therefore, even if the apparatus (in this case, the quantum computer) is ideal, there will be statistical fluctuations in the result due to limited number of trials in the experiment. For example, if we conduct $N$ trials from which we estimate the value of $P_i = \Mod{\braket{i|\psi}}^2$, the observed probability (the relative frequency of the outcome) will have a Gaussian distribution around the true probability value, with the mean of the distribution being the true probability value $P_i$ and the standard deviation being $\sqrt{P_i (1-P_i)/N_i}$. The fluctuation in the observed probability $\Delta P_i$ with about 95\% confidence interval is given by
\begin{align}
 \Delta P_i =& \pm 1.96 \sqrt{\frac{P_i(1-P_i)}{N}}
\end{align}
\begin{figure}
 \centering
 \includegraphics[width=0.5\textwidth]{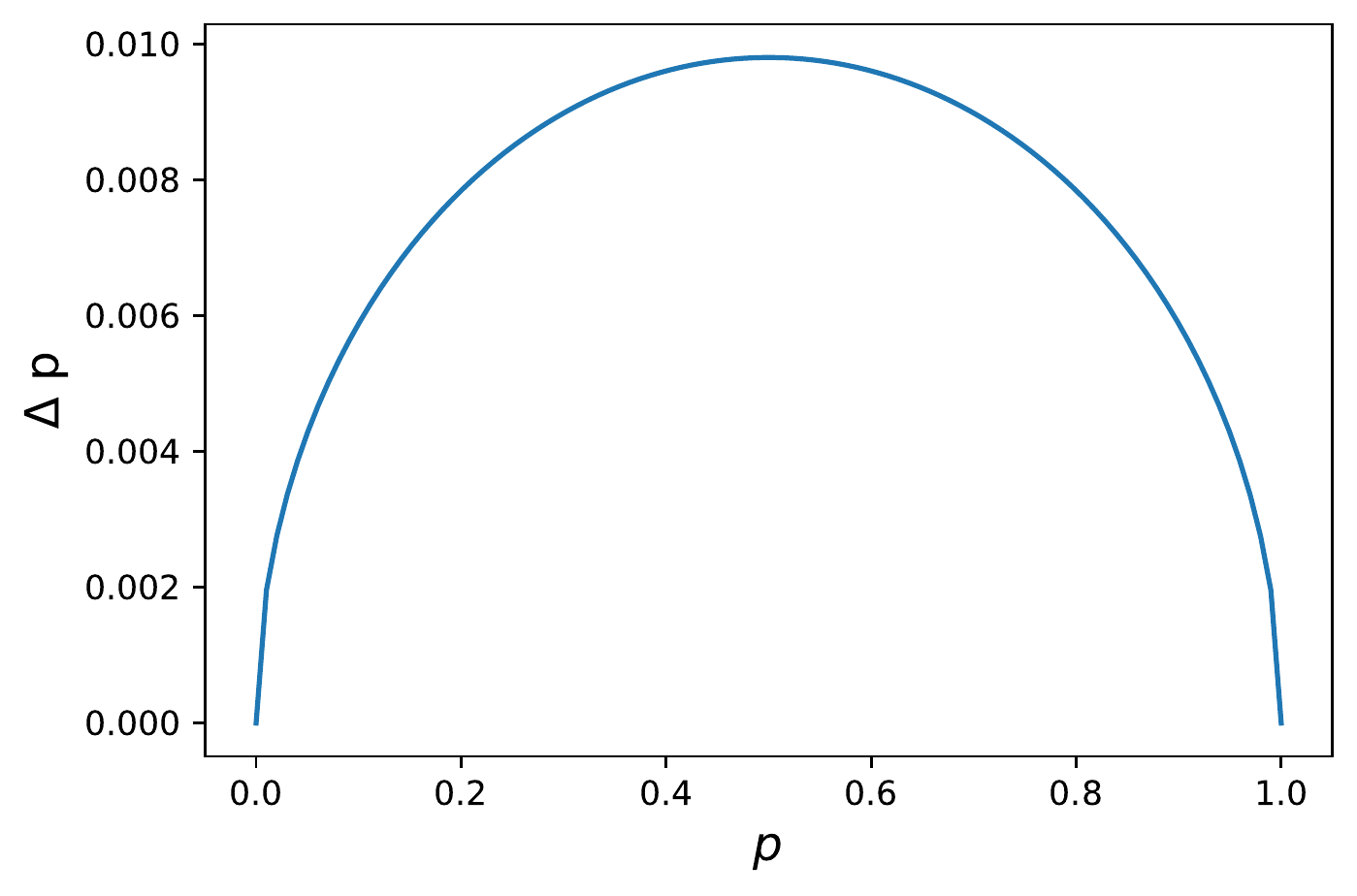}
 \caption{Plot of error in the relative frequency of outcome as a function of the true probability. The number of trials is $10^4$.}
 \label{fig:error_probs}
\end{figure}
Figure \ref{fig:error_probs} shows the error in the relative frequency of outcome as function of the true probability of that outcome, for $N=10^4$. And in terms of projective measurement outcomes,
\begin{align}
 \Delta P_i =& \pm 1.96 \sqrt{\frac{\Mod{\braket{0|\psi}}^2 \Mod{\braket{1|\psi}}^2}{N}}
\end{align}
This is the maximum fluctuation of the measurement frequency about the true probability with 95\% confidence interval.
This fluctuation in the observed probability will propagate to the quantities that we calculate using the probability values. To calculate the error in gamma values, we differentiate equation \eqref{eq:gamma12i},
\begin{align}
 \ln\gamma_{12}^{(i)} =& \ln\left(N^2 P_{12}^{(i)} - p^2 P_{1}^{(i)} - (1-p^2) P_{2}^{(i)}\right) \nonumber \\
 &- \frac{1}{2}\ln P_{1}^{(i)} - \frac{1}{2}\ln P_2^{(i)} - \ln\left(2 p \sqrt{1-p^2} \right)\\
 \Delta\gamma_{12}^{(i)} =& \gamma_{12}^{(i)}\left(\frac{N^2 \Delta P_{12}^{(i)} - p^2 \Delta P_{1}^{(i)} - (1-p^2) \Delta P_{2}^{(i)}}{N^2 P_{12}^{(i)} - p^2 P_{1}^{(i)} - (1-p^2) P_{2}^{(i)}} \right. \nonumber \\
 &\left.- \frac{1}{2}\frac{\Delta P_1^{(i)}}{P_1^{(i)}} - \frac{1}{2}\frac{\Delta P_2^{(i)}}{P_2^{(i)}} \right)
\end{align}
where $P_\alpha^i = \Mod{\braket{i|\psi_\alpha}}^2$. The maximum error in the value will be 
\begin{align}
 \Delta\gamma_{12}^{(i)} =& \Mod{\gamma_{12}^{(i)}}\left(\frac{N^2 \Mod{\Delta P_{12}^{(i)}} + p^2 \Mod{\Delta P_{1}^{(i)}} + (1-p^2) \Mod{\Delta P_{2}^{(i)}}}{N^2 P_{12}^{(i)} - p^2 P_{1}^{(i)} - (1-p^2) P_{2}^{(i)}} \right.\nonumber \\
 &\left.+ \frac{1}{2}\frac{\Mod{\Delta P_1^{(i)}}}{P_1^{(i)}} + \frac{1}{2}\frac{\Mod{\Delta P_2^{(i)}}}{P_2^{(i)}} \right)
\end{align}
Because we are adding two terms to calculate the value of $\gamma_{12}$, the total error is 
\begin{align}
 \Delta \gamma_{12} =& \frac{1}{2} \left(\Delta\gamma_{12}^{(0)} + \Delta\gamma_{12}^{(1)}\right)
\end{align}
This is the fluctuation in the value of $\gamma$ due to the statistical fluctuation in the observed measurement probability, therefore this error will be present even if the quantum computer is ideal.
Finally, the error in the value of $F$ is 
\begin{align}
 \Delta F =& 2 \Delta\gamma_{12} \gamma_{12} + 2 \Delta\gamma_{23} \gamma_{23} + 2 \Delta\gamma_{31} \gamma_{31} \nonumber \\
 &- 2 \Delta\gamma_{12} \gamma_{23} \gamma_{31} - 2 \gamma_{12} \Delta\gamma_{23} \gamma_{31}  - 2 \gamma_{12} \gamma_{23} \Delta\gamma_{31}  \\
 \Delta F =& 2 \Delta\gamma_{12} \left(\gamma_{12} - \gamma_{23} \gamma_{31}\right) + 2 \Delta\gamma_{23} \left(\gamma_{23} - \gamma_{12} \gamma_{31}\right) \nonumber \\
 &+ 2 \Delta\gamma_{31} \left(\gamma_{31} - \gamma_{12} \gamma_{23}\right)
\end{align}
The maximum value of $\Delta F$ can be found by checking for different signs of $\Delta \gamma$.
\end{document}